\documentclass[preprint,12pt]{elsarticle}
\usepackage{amssymb}
\usepackage{amsmath}
\usepackage{mathtools, nccmath}
\usepackage{graphicx}
\usepackage[T1]{fontenc}
\usepackage[polish,english]{babel}
\usepackage[utf8]{inputenc}\usepackage{mathtools}
\usepackage{dsfont}
\usepackage{amsthm}
\newtheorem{definition}{Definition}
\newtheorem{example}{Example}

\usepackage{array}
\usepackage{enumitem}
\usepackage{pdflscape}
\usepackage{hyperref}
\newcolumntype{M}[1]{>{\centering\arraybackslash}m{#1}}
\usepackage{multirow}
\usepackage{color,soul}
\usepackage{pifont}% http://ctan.org/pkg/pifont
\usepackage{makecell}

\usepackage{ltablex}
\usepackage{longtable}
\usepackage{tabularx}
\usepackage{multirow}
\usepackage{enumitem}
\newcolumntype{C}{>{\Centering\arraybackslash}X} % centered "X" column
\usepackage[width=1.0\textwidth]{caption}
\usepackage{makecell}
\usepackage{verbatim}% 

\journal{Theoretical Computer Science}

\begin{document}

\begin{frontmatter}

%% Title, authors and addresses

%% use the tnoteref command within \title for footnotes;
%% use the tnotetext command for theassociated footnote;
%% use the fnref command within \author or \address for footnotes;
%% use the fntext command for theassociated footnote;
%% use the corref command within \author for corresponding author footnotes;
%% use the cortext command for theassociated footnote;
%% use the ead command for the email address,
%% and the form \ead[url] for the home page:
%% \title{Title\tnoteref{label1}}
%% \tnotetext[label1]{}
%% \author{Name\corref{cor1}\fnref{label2}}
%% \ead{email address}
%% \ead[url]{home page}
%% \fntext[label2]{}
%% \cortext[cor1]{}
%% \affiliation{organization={},
%%             addressline={},
%%             city={},
%%             postcode={},
%%             state={},
%%             country={}}
%% \fntext[label3]{}

\title{QEGS: A Mathematica Package for the Analysis of Quantum Extended Games}

%% use optional labels to link authors explicitly to addresses:
%% \author[label1,label2]{}
%% \affiliation[label1]{organization={},
%%             addressline={},
%%             city={},
%%             postcode={},
%%             state={},
%%             country={}}
%%
%% \affiliation[label2]{organization={},
%%             addressline={},
%%             city={},
%%             postcode={},
%%             state={},
%%             country={}}

\author[inst1]{Krzysztof Grzanka}
\author[inst1]{Anna Gorczyca-Goraj}
\author[inst2]{Piotr Frąckiewicz}
\author[inst1]{Marek Szopa\corref{cor1}}
\ead{marek.szopa@uekat.pl}

\cortext[cor1]{Corresponding Author}
\affiliation[inst1]{organization={Department of Operations Research, University of Economics in Katowice},%Department and Organization
            addressline={ul. Bogucicka 3}, 
            city={Katowice},
            postcode={40-287}, 
            country={Poland}}

\affiliation[inst2]{organization={Institute of Exact and Technical Sciences, Pomeranian University in Słupsk},%Department and Organization
            addressline={ul. Arciszewskiego 22A}, 
            city={Słupsk},
            postcode={76-200}, 
            country={Poland}}

\begin{abstract}
Quantum games have attracted much attention in recent years due to their ability to solve decision-making dilemmas. The aim of this study is to extend previous work on quantum games by introducing a Mathematica package QEGS (\textit{Quantum Extension Game Solver}) dedicated to the study of quantum extensions of classical $2\times2$ games based on the EWL scheme. The package generates all possible game extensions with one or two unitary strategies, which are invariant with respect to isomorphic transformations of the initial games. The package includes a number of functions to study these extensions, such as determining their Nash equilibria in pure strategies, eliminating dominated strategies, or computing maximin strategies. Independently of quantum extensions, these functions can also be used to analyze classical games. Reporting to a pdf is available. The discussion includes an outline of future research directions, such as the exploration of mixed-strategy Nash equilibria and potential real-world applications in fields like quantum computing and secure communications.

\end{abstract}

%%Graphical abstract
%%\begin{graphicalabstract}
%%\includegraphics{grabs}
%%\end{graphicalabstract}

%%Research highlights
\begin{highlights}
\item QEGS: A Mathematica package for examining characteristics of quantum extended games.
\item Implements systematic quantum game extensions based on the Eisert–Wilkens–Lewenstein (EWL) scheme.
\item Evaluates strategic properties including pure strategy Nash equilibria, dominance, and maximin strategies.
\item Provides computational tools facilitating exploration of classical and quantum strategic scenarios.
\item Bridges theoretical developments with practical applications in quantum decision sciences and computing.
\end{highlights}

\begin{keyword}
%% keywords here, in the form: keyword \sep keyword
game theory \sep quantum game \sep Mathematica package \sep Eisert-Wilkens-Lewenstein scheme \sep   Nash equilibrium
%% PACS codes here, in the form: \PACS code \sep code
\PACS 02.50.Le \sep 03.67.Ac
%% MSC codes here, in the form: \MSC code \sep code
%% or \MSC[2008] code \sep code (2000 is the default)
\MSC 81-08 \sep 91-08 \sep 91A05 \sep 91A80
\end{keyword}

\end{frontmatter}

\newpage

{\bf PROGRAM SUMMARY}

\begin{small}
\noindent
{\em Authors: } K. Grzanka, A. Gorczyca-Goraj, P. Frąckiewicz, M. Szopa          \\
{\em Developer's repository:} \href{https://github.com/k-grzanka/QEGS}{https://github.com/k-grzanka/QEGS}  \\
{\em Licensing provisions:} GNU Public License v3 \\
{\em Programming language:} \texttt{Wolfram Mathematica 10} or higher (Wolfram Language) \\
{\em Nature of problem:} Creating the quantum extended games based on the EWL scheme. Examination of these extensions and other bimatrix games' properties.\\
{\em Restrictions:} Depending on the complexity of the matrix, limited by memory and CPU.\\
{\em References:}
\begin{enumerate}
    \item[1] \href{https://www.wolfram.com/mathematica}{https://www.wolfram.com/mathematica}, commercial algebraic software
\end{enumerate}
\end{small}

\newpage

\section{Introduction}
\label{sec:introduction}
Quantum game theory, as an interdisciplinary field, combines the principles of classical game theory with the concepts of quantum mechanics to analyze strategic interactions in scenarios enriched with quantum phenomena. Since its inception, this field has attracted attention for its potential to provide deeper insights into decision-making processes and to explore novel strategies unavailable in classical frameworks. The seminal work by Eisert, Wilkens, and Lewenstein (EWL) introduced a formal quantization scheme for classical games, which has since become a cornerstone of quantum game theory \cite{eisert_quantum_1999}. Building on these foundations, researchers have demonstrated how quantum strategies can impact classical dilemmas, including the well-known Prisoner’s Dilemma (PD), by shifting Nash equilibria toward Pareto-optimal solutions \cite{flitney_introduction_2002,meyer_quantum_1999}.

In classical game theory, the Nash equilibrium (NE) is a central solution concept that identifies strategy profiles where no player can improve their payoff by unilaterally changing their strategy. However, in classical games like the PD, NE often fails to align with socially optimal outcomes, as illustrated by mutual defection being the dominant strategy \cite{nash_non-cooperative_1951}. Quantum extensions of such games, particularly within the EWL framework, introduce additional unitary strategies that expand the players' strategic spaces. This enables equilibria that are not only more varied but also more aligned with Pareto-optimal outcomes \cite{eisert_quantum_1999,marinatto_quantum_2000}.

Despite significant progress in understanding quantum games, challenges remain. One prominent issue is the complexity of deriving and analyzing NE in quantum settings, especially when extensions involve multiple quantum strategies. This complexity arises from the interplay between the classical payoff structure and the parameters governing quantum strategies defined by unitary operators characterized by angles and phases \cite{du_entanglement_2002,benjamin_multiplayer_2001}. Addressing these challenges requires both theoretical advancements and practical tools to simplify the investigation of quantum games.

The aim of this study is to extend previous work on quantum game theory by introducing a dedicated Mathematica package designed to investigate the existence and properties of NE in quantum extended games. This package builds upon the methodologies developed in earlier studies on quantum extensions of the $2\times2$ classical games \cite{frackiewicz_permissible_2024-1,frackiewicz_permissible_2024}. It offers a versatile platform for exploring game extensions with one or two quantum strategies, which are invariant with respect to isomorphic transformations of the initial game. Key features of the package include:
\begin{itemize}
    \item Comprehensive Analysis of Extensions. The package enables users to define classical games as inputs and generate quantum extensions with one or two quantum strategies. 
    \item Optimization Features. To streamline strategic analysis, the package provides functionalities to calculate NE in pure strategies, analyze maximin strategies and  supports the elimination of dominated strategies, allowing users to focus on optimal decision-making pathways.
    \item Classical games. The calculation of NE in pure strategies, the elimination of dominated strategies and the determination of maximin strategies can also be applied to any classical game.
    \item Customizable Output. Results can be reported in PDF format, facilitating detailed documentation and dissemination of findings.
\end{itemize}
This study leverages results from prior research on quantum extensions of games like the Prisoner’s Dilemma to benchmark the Mathematica package and validate its functionalities. By illustrating its application to commonly studied games, the paper provides practical examples of how quantum game theory can be utilized to address decision-making dilemmas. Moreover, the discussion includes an outline of future research directions, such as the exploration of mixed-strategy NE and potential real-world applications in fields like quantum computing and secure communications.

In summary, this paper contributes to the growing body of quantum game theory by offering a computational tool to simplify the analysis of NE in quantum games. It aims to bridge the gap between theoretical advancements and practical applications, making quantum game theory more accessible to researchers and practitioners alike.

\section{Preliminaries}
\begin{definition}
A bimatrix game is a two-player game in which each player has a finite set of strategies, and the outcomes are determined by two payoff matrices, one for each player. A bimatrix game can be represented by 
\begin{equation}\label{generalbimatrix}
\Delta = 
\begin{pmatrix}
(\Delta^1_{11}, \Delta^2_{11}) & (\Delta^1_{12}, \Delta^2_{12}) & \cdots &(\Delta^1_{1m}, \Delta^2_{1m}) \\ 
(\Delta^1_{21}, \Delta^2_{21}) & (\Delta^1_{22}, \Delta^2_{22}) & \cdots &(\Delta^1_{2m}, \Delta^2_{2m}) \\ 
\vdots & \vdots & \ddots & \vdots \\ 
(\Delta^1_{n1}, \Delta^2_{n1}) & (\Delta^1_{n2}, \Delta^2_{n2}) & \cdots &(\Delta^1_{nm}, \Delta^2_{nm})
\end{pmatrix} = (\Delta^1, \Delta^2).
\end{equation}
\end{definition}
%%%%%%%%%%%%%%%%%%%%%%%%%%%%%%%%%%%%%%%
The interpretation of such a notation is that player 1 (the row player) chooses row $i\in \{1, \dots, n\}$ and player 2 (the column player) chooses column $j\in \{1, \dots, m\}$. Rows $i$ and columns 
$j$ are generally referred to as the players' strategies. The combination of player 1 using strategy $i$ and player 2 using strategy $j$ will be represented as the ordered pair $(i,j)$ and referred to as a strategy profile. As the result of the game, player 1 receives a payoff $\Delta^1_{ij}$ and player 2 receives $\Delta^2_{ij}$. 
%%%%%%%%%%%%%%%%%%%%%%%%%%%%%%%%%%%%%%%
\begin{definition}
A bimatrix game $(\Delta^1, \Delta^2)$ is said to be symmetric if $\Delta^2 = (\Delta^1)^{T}$.
\end{definition}
Since, in a symmetric game, the payoff matrix of one player is determined by the payoff matrix of the other player, we can simplify the notation by expressing the game using only the payoff matrix $\Delta^1$ of Player 1.
\begin{example}\label{Example_1}
A popular example of a symmetric game is the Prisoner's Dilemma, for which the bimatrix form can be expressed as follows:
\begin{equation}\label{PDg}
\begin{pmatrix}
(3,3) & (0,5) \\ 
(5,0) & (1,1)
\end{pmatrix}.
\end{equation}
If it is explicitly stated that the game under consideration is symmetric, it is sufficient to provide the payoff matrix of player 1, which uniquely determines the payoff matrix of player 2.
\end{example}
%%%%%%%%%%%%%%%%%%%%%%%%%%%%%%%%%%%%%%%%%%%%
Smaller games are preferable for computational efficiency. In some cases, a bimatrix game can be simplified by removing rows or columns (i.e., strategies) that will never be chosen, as there is always a superior alternative available. This process is known as dominance elimination. Before analyzing a game, it is beneficial to check for dominance, as it can help reduce the matrix size and simplify calculations.
\begin{definition}
Let $(\Delta^1, \Delta^2)$ be an $n\times m$ bimatrix game. The pure strategy $i\in \{1, \dots, n\}$ of player 1 is strictly dominated if there exists another strategy $i'\in \{1,\dots, n\}$ such that $\Delta^1_{i'j} > \Delta^1_{ij}$ for every strategy $j\in \{1, \dots, m\}$ of player 2. Similarly, the pure strategy $j\in \{1, \dots, m\}$ of player 2 is strictly dominated if there exists another strategy $j'\in \{1,\dots, m\}$ such that $\Delta^2_{ij'} > \Delta^2_{ij}$ for every strategy $i\in \{1, \dots, n\}$ of player 1.
\end{definition}
\begin{example}
Let us consider game~(\ref{PDg}). Player 1's first strategy is strictly dominated by her second strategy. Indeed, 
\begin{equation}
\Delta^1_{21} > \Delta^1_{11} \quad \text{and} \quad \Delta^1_{22} > \Delta^1_{12}.
\end{equation}
Analogously, one can check that player 2's first strategy is strictly dominated in (\ref{PDg}).
\end{example}
%%%%%%%%%%%%%%%%%%%%%%%%%%%%%%%%%%%%%%%%%%%%
One of the fundamental solution concepts in game theory is the NE, which can be expressed in terms of bimatrix games as follows: 
\begin{definition}\label{NEdef}
A strategy profile $(i^*, j^*)$ is a (pure) NE in $(\Delta^1, \Delta^2)$ if $\Delta^1_{i^*j^*} \geq \Delta^1_{ij^*}$ for every $i\in \{1, \dots n\}$ and $\Delta^2_{i^*j^*} \geq \Delta^2_{i^*j}$ for every $j\in \{1, \dots m\}$.
\end{definition}
A NE represents a stability property in which no player has an incentive to unilaterally deviate from their chosen strategy.
\begin{example}
According to Definition~\ref{NEdef}, the unique NE in~\ref{PDg} is the strategy profile $(2,2)$, i.e., the profile in which the players choose their second row and second column, respectively. Indeed, 
\begin{equation}
\Delta^1_{22} \geq \Delta^1_{12} \quad \text{and} \quad \Delta^2_{22} \geq \Delta^2_{21}.
\end{equation}
\end{example}
%%%%%%%%%%%%%%%%%%%%%%%%%%%%%%%%%%%%%%%%%%%%%

The concept of a security strategy that guarantees the best outcome without relying on what the opponent will do, assuming the most pessimistic scenario.
\begin{definition}
A (pure) strategy $i\in \{1,\dots, n\}$ is a maximin strategy of player 1 in a bimatrix game (\ref{generalbimatrix}) if 
\begin{equation}
    \min_{j\in \{1, \dots m\}}\Delta^1_{ij} \geq \min_{j\in \{1, \dots m\}}\Delta^1_{i'j} ~~\text{for all}~~ i' \in \{1, \dots, n\}.
\end{equation}
A (pure) strategy $j\in \{1,\dots, m\}$ is a maximin strategy of player 2 in a bimatrix game ... if
\begin{equation}
    \min_{i\in \{1, \dots n\}}\Delta^2_{ij} \geq \min_{i\in \{1, \dots n\}}\Delta^2_{ij'} ~~\text{for all}~~ j' \in \{1, \dots, m\}.
\end{equation}
\end{definition}
\begin{example}\label{Example_3}
Let us consider the following game:
\begin{equation}\label{game_2_3}
\bordermatrix{& L & M & R \cr
T & (3,1) & (2, 3) & (2,0) \cr
B & (-100,1) & (-100, 2) & (3,3)
}
\end{equation}
Let us note that the game has a unique pure NE $(B,R)$. However, due to the high risk for player 1 of receiving -100 instead of 3, the most likely outcome of the game is the strategy profile predicted by the concept of maximin strategies. The maximin strategy of player 1 is $T$, while the maximin strategy of player 2 is $M$. Thus, the outcome resulting from playing these strategies is $(T,M)$ with a payoff of $(2,3)$. 
\end{example}
%%%%%%%%%%%%%%%%%%%%%%%%%%%%%%%%%%%%%%%%%%%%%%%
The Eisert-Wilkens-Lewenstein (EWL) scheme is one of the fundamental approaches to modeling quantum games, incorporating quantum mechanics into classical game theory. Proposed in 1999 by Jens Eisert, Martin Wilkens and Maciej Lewenstein \cite{eisert_quantum_1999}, this scheme extends $2\times 2$ bimatrix games (game~(\ref{generalbimatrix})  for $n=m =2$) by employing quantum logic gates and entanglement, leading to new strategies and outcomes that are unattainable within classical frameworks.
In the EWL model, players operate within a Hilbert space, where their strategies are represented by unitary operators
\begin{equation}\label{qstrategy}
U_{i}(\theta_{i}, \alpha_{i}, \beta_{i}) = \begin{pmatrix}
e^{i\alpha_{i}}\cos\frac{\theta_{i}}{2} & ie^{i\beta_{i}}\sin\frac{\theta_{i}}{2} \\ 
ie^{-i\beta_{i}}\sin\frac{\theta_{i}}{2} & e^{-i\alpha_{i}}\cos\frac{\theta_{i}}{2}
\end{pmatrix}, \quad \theta_{i} \in [0,\pi] ~\text{and}~ \alpha_{i}, \beta_{i} \in [0,2\pi),
\end{equation}
acting on qubits. The game begins with the preparation of an initial state in the form of a maximally entangled state, after which each player applies a chosen unitary transformation (\ref{qstrategy}) on his own qubit of this state. The final outcome is then determined through measurement in the computational basis. A suitably defined payoff function allows for analyzing the impact of quantum mechanics on game results, revealing potential advantages over classical strategies, including the existence of quantum NE, which may have no counterparts in classical game theory.
A full description of the EWL scheme is presented in \cite{frackiewicz_permissible_2024-1}. For the purposes of this article, we recall the derived formula for the payoff functions:
\begin{align}\label{generalEWLpayoff}
    &u(U_{1}(\theta_{1}, \alpha_{1}, \beta_{1}), U_{2}(\theta_{2}, \alpha_{2}, \beta_{2})) \nonumber\\ &\quad = (\Delta^1_{11}, \Delta^2_{11})\left(\cos(\alpha_{1} + \alpha_{2})\cos\frac{\theta_{1}}{2}\cos\frac{\theta_{2}}{2} + \sin(\beta_{1} + \beta_{2})\sin\frac{\theta_{1}}{2}\sin\frac{\theta_{2}}{2}\right)^2 \nonumber\\
    &\quad + (\Delta^1_{12}, \Delta^2_{12})\left(\cos(\alpha_{1} - \beta_{2})\cos\frac{\theta_{1}}{2}\sin\frac{\theta_{2}}{2} + \sin(\alpha_{2} - \beta_{1})\sin\frac{\theta_{1}}{2}\cos\frac{\theta_{2}}{2}\right)^2 \nonumber \\ 
    &\quad + (\Delta^1_{21}, \Delta^2_{21})\left(\sin(\alpha_{1} - \beta_{2})\cos\frac{\theta_{1}}{2}\sin\frac{\theta_{2}}{2} + \cos(\alpha_{2} - \beta_{1})\sin\frac{\theta_{1}}{2}\cos\frac{\theta_{2}}{2}\right)^2 \nonumber \\ 
    &\quad + (\Delta^1_{22}, \Delta^2_{22})\left(\sin(\alpha_{1} + \alpha_{2})\cos\frac{\theta_{1}}{2}\cos\frac{\theta_{2}}{2} - \cos(\beta_{1} + \beta_{2})\sin\frac{\theta_{1}}{2}\sin\frac{\theta_{2}}{2}\right)^2.
\end{align}
%%%%%%%%%%%%%%%%%%%%%%%%%%%%%%%%%%%%%%%%%%%%%%%
Using (\ref{generalEWLpayoff}), we can determine a finite extension of the classical $2\times 2$ game with quantum strategies. If the extension consists of adding a single quantum strategy, then the players' strategy sets are:
\begin{equation}\label{3set}
\{U(0,0,0), U(\pi,0,0), U(\theta, \alpha, \beta)\}.
\end{equation}
The operator $U(0,0,0)$ is the identity matrix, which we will denote by $I$. The operator $U(\pi, 0,0)$ is $iX$ - the Pauli matrix X multiplied by the imaginary unit. These strategies can be identified with classical strategies. The operator $U(\theta, \alpha, \beta) \equiv U$ is an arbitrary but fixed unitary strategy. 

From~(\ref{generalEWLpayoff}) one can verify that 
\begin{equation}
\begin{array}{ll}
u(I,I) = (\Delta^1_{11}, \Delta^2_{11}), & u(I,iX) = (\Delta^1_{12}, \Delta^2_{12}), \\
u(iX, I) = (\Delta^1_{21}, \Delta^2_{21}), & u(iX, iX) = (\Delta^1_{22}, \Delta^2_{22}).
\end{array}
\end{equation}
The payoff profiles resulting from playing the strategy profiles that include strategy $U$ are
\begin{multline}
u(U, I) = (\Delta^1_{11}, \Delta^2_{11})\cos^2\alpha\cos^2\frac{\theta}{2} + (\Delta^1_{12}, \Delta^2_{12})\sin^2\beta\sin^2\frac{\theta}{2}\\ + (\Delta^1_{21}, \Delta^2_{21})\cos^2\beta\sin^2\frac{\theta}{2} + (\Delta^1_{22}, \Delta^2_{22})\sin^2\alpha\cos^2\frac{\theta}{2},
\end{multline}
%%%%%%%%%%%%%%%%%%%%%%%%%%%%
\begin{multline}
u(U, iX) =  (\Delta^1_{11}, \Delta^2_{11})\sin^2\beta\sin^2\frac{\theta}{2} + (\Delta^1_{12}, \Delta^2_{12})\cos^2\alpha\cos^2\frac{\theta}{2}\\ + (\Delta^1_{21}, \Delta^2_{21})\sin^2\alpha\cos^2\frac{\theta}{2} + (\Delta^1_{22}, \Delta^2_{22})\cos^2\beta\sin^2\frac{\theta}{2},
\end{multline}
%%%%%%%%%%%%%%%%%%%%%%%%%%%%%%%%%%%%%%
\begin{multline}
u(I, U) = (\Delta^1_{11}, \Delta^2_{11})\cos^2\alpha\cos^2\frac{\theta}{2} + (\Delta^1_{12}, \Delta^2_{12})\cos^2\beta\sin^2\frac{\theta}{2} \\ + (\Delta^1_{21}, \Delta^2_{21})\sin^2\beta\sin^2\frac{\theta}{2} + (\Delta^1_{22}, \Delta^2_{22})\sin^2\alpha\cos^2\frac{\theta}{2},
\end{multline}
%%%%%%%%%%%%%%%%%%%%%%%%%%%%%%%%%%%%%%%
\begin{multline}
u(iX, U) = (\Delta^1_{11}, \Delta^2_{11})\sin^2\beta\sin^2\frac{\theta}{2} + (\Delta^1_{12}, \Delta^2_{12})\sin^2\alpha\cos^2\frac{\theta}{2}\\ + (\Delta^1_{21}, \Delta^2_{21})\cos^2\alpha\cos^2\frac{\theta}{2} + (\Delta^1_{22}, \Delta^2_{22})\cos^2\beta\sin^2\frac{\theta}{2},
%%%%%%%%%%%%%%%%%%%%%%%%%%%%%%%%%%%%%%%%
\end{multline}
\begin{align}\label{wyplatauu}
u(U, U) &= (\Delta^1_{11}, \Delta^2_{11})\left(\cos(2\alpha)\cos^2\frac{\theta}{2} + \sin(2\beta)\sin^2\frac{\theta}{2}\right)^2 \nonumber \\ 
&\quad + 
%%%%%%%%%%%%%%%%%%%%%%%%%%%%%%%%%%%%%
\frac{1}{4}((\Delta^1_{12}, \Delta^2_{12})+(\Delta^1_{21}, \Delta^2_{21}))\left(\cos(\alpha-\beta) + \sin(\alpha-\beta)\right)^2\sin^2\theta \nonumber \\
&\quad +
%%%%%%%%%%%%%%%%%%%%%%%%%%%%%%%%%%%%%%%%%
(\Delta^1_{22}, \Delta^2_{22})\left(\sin(2\alpha)\cos^2\frac{\theta}{2} - \cos(2\beta)\sin^2\frac{\theta}{2}\right)^2.
\end{align}
Then a finite $3\times 3$ quantum extension of the game is 
\begin{equation}\label{ext_3x3}
\bordermatrix{& I & iX & U \cr
I & (\Delta^1_{11}, \Delta^2_{11}) & (\Delta^1_{12}, \Delta^2_{12}) & u(I,U) \cr 
iX & (\Delta^1_{21}, \Delta^2_{21}) & (\Delta^1_{22}, \Delta^2_{22}) & u(iX, U) \cr
U & u(U,I) & u(U, iX) & u(U,U)
}.
\end{equation}
It is worth noting that the choice of the unitary strategy $U$ is not arbitrary and must be based on certain principles, which will be addressed in the next section.
\begin{example}
Let us consider~(\ref{PDg}) and strategy set~(\ref{3set}) in the form of 
\begin{equation}
\left\{U(0,0,0), U(\pi,0,0), U\left(\frac{\pi}{3}, \frac{\pi}{2}, \pi\right)\right\}. 
\end{equation}
The payoff pairs corresponding to strategy profiles in which at least one player plays strategy U are
\begin{equation}
\begin{aligned}
&u(I, U) = \left(\frac{3}{4}, 2\right), u(U, I) = \left(2, \frac{3}{4}\right), u(iX, U) = \left(\frac{3}{4}, 2\right), \\
&u(I, U) = \left(\frac{3}{4}, 2\right), u(I, U) = \left(\frac{3}{4}, 2\right).
\end{aligned}
\end{equation}
As a result, the extension of the bimatrix (2) by one quantum strategy is as follows:
\begin{equation}\label{newgameex}
\bordermatrix{& I & iX & U \cr
I & (3, 3) & (0, 5) & \left(\frac{3}{4}, 2\right) \cr 
iX & (5, 0) & (1, 1) & \left(\frac{1}{4},4\right) \cr
U & \left(2, \frac{3}{4}\right) & \left(4, \frac{1}{4}\right) & \left(\frac{43}{16}, \frac{43}{16}\right)
}.
\end{equation}
Adding the strategy $U$ to (\ref{PDg}) completely changes the course of the game. The strategy profile $(iX, iX)$ is no longer a NE, and the unique NNE in game (\ref{newgameex}) is now $(U,U)$ with a payoff of 43/16 for each player. It can thus be concluded that players who are parties to the problem described by the PD, by making available to them an additional quantum strategy, can expect a much better outcome of their strategic interaction than if they had only classical operations at their disposal. It should be noted that although the extended game is based on quantum operations, when the final states of the qubits are read, it produces a completely classical result, which is the basis for deciding which of the pure strategies of the classical game (e.g. cooperation or defection) to choose.
\end{example}
\vspace{12pt}
%%%%%%%%%%%%%%%%%%%%%%%%%%%%%%%%%%%%%%%%%%%%%%%%%%%%%%%%%%%%%%%%%%%%%%%%%%%%%%%%%%%%%%%%%%%%%%%%%%%%%%%%%%%%%%%%%%%%%%%%%%%%%%%%%%%%%%
\section{Permissible quantum extensions by one or two unitary strategies}
\label{sec:four-strategy_quantum_extensions}

In the current chapter, we discuss the necessary conditions that quantum extensions of the classical game must satisfy. The QEGS package starts with arbitrary $2\times2$ classical game
\begin{equation}\label{classical2x2}
\Gamma=\begin{pmatrix}
         \Delta_{11} & \Delta_{12}\\
         \Delta_{21} & \Delta_{22}
\end{pmatrix}, \text{ where  } \Delta_{ij}=(\Delta^1_{ij},\Delta^2_{ij}).
\end{equation}
The extension of the classical game is the addition of one (\ref{ext_3x3}) or two quantum strategies to the two classical strategies. Quantum strategies involve players employing quantum operations (unitary transformations) on their quantum bits (qubits). As demonstrated in \cite{frackiewicz_permissible_2024-1,frackiewicz_permissible_2024}, there exist specific permissible methods to extend classical games by utilizing one or two unitary operations along with two strategies that replicate those of the classical game. The allowed extensions must be invariant with respect to the isomorphic transformation of the initial game. These permissible extensions are categorized into different classes based on parameters that define additional unitary strategies. Each class involves certain matrix manipulations and transformations that are consistent with the game's isomorphic transformations. The focus on NE implies studying how these quantum strategies alter the outcomes and stability of the equilibrium compared to the classical scenario.

The space of all possible game states is richer for a quantum game than for a classical game. Classic players have only two so-called pure states (e.g., in the case of the Prisoner's Dilemma (\ref{PDg}) it is cooperation and defection) and mixed states, i.e. probabilistic mixtures of pure states. In the quantum extension, each player has a qubit, which is a superposition of the two states with complex coefficients, forming the so-called Bloch sphere. The qubits of both players are entangled with each other, which accounts for the quantum correlations of their strategies, which are not available in the classical game. In the $3\times3$ extension, each player has three strategies - unitary operators acting on their qubits. Identity operator $I$ - corresponds to the passive strategy, of not changing the state of the qubit. Pauli matrix multiplied by the imaginary unit $iX$ - corresponds to the classical strategy of swapping game states. Both these operators accurately describe the operations possible for classical players. The third, unitary operator $U$ (\ref{qstrategy}) is precisely defined by assuming that the quantum expansion is invariant to isomorphic transformations of the classical game. This presumption is essential for a quantum game to be regarded as an extension of a specific classical game \cite{frackiewicz_permissible_2024-1}. It also determines the payoff matrix (\ref{ext_3x3}), calculated accordingly to the $U$ operator. Consequently, the quantum expansion matrix can be expressed by: 

\begin{equation}\label{general3x3}
\bordermatrix{ & I & iX & U \cr I & \Delta_{11} & \Delta_{12} & \Delta_{13} \cr 
iX & \Delta_{21} & \Delta_{22} & \Delta_{23} \cr 
U & \Delta_{31} & \Delta_{32} & \Delta_{33} 
}. 
\end{equation}
As it was shown in \cite{frackiewicz_permissible_2024-1}, there are only three permissible types of $3\times 3$ extensions: 
\begin{equation}\label{A0}
A_0=
\begin{pmatrix} 
\Delta_{11} & \Delta_{12} & \frac{\Delta_{11}+\Delta_{12}}{2} \cr 
\Delta_{21} & \Delta_{22} & \frac{\Delta_{21}+\Delta_{22}}{2} \cr 
\frac{\Delta_{11}+\Delta_{21}}{2} & \frac{\Delta_{12}+\Delta_{22}}{2} & \frac{\Delta_{11}+\Delta_{12} + \Delta_{21} + \Delta_{22}}{4}
\end{pmatrix},
\end{equation}
where the corresponding $U=U(\frac{\pi}{2},\alpha, \beta)$, and $\alpha, \beta \in \{0,\pi\}$.
The second class is
\begin{equation}\label{B0}
B_0=
\begin{pmatrix}
\Delta_{11} & \Delta_{12} & \frac{\Delta_{21}+\Delta_{22}}{2} \cr 
\Delta_{21} & \Delta_{22} & \frac{\Delta_{11}+\Delta_{12}}{2} \cr 
\frac{\Delta_{12}+\Delta_{22}}{2} & \frac{\Delta_{11}+\Delta_{21}}{2} & \frac{\Delta_{11}+\Delta_{12} + \Delta_{21} + \Delta_{22}}{4} 
\end{pmatrix},
\end{equation}
where $U=U(\frac{\pi}{2},\alpha, \beta)$, and $\alpha, \beta \in \{\frac{\pi}{2},\frac{3\pi}{2}\}$. And the third class is
\begin{equation}\label{C_0}
C_0=
\begin{pmatrix}
\Delta_{11} & \Delta_{12} & \frac{\Delta_{11}+\Delta_{12} + \Delta_{21} + \Delta_{22}}{4} \cr 
\Delta_{21} & \Delta_{22} & \frac{\Delta_{11}+\Delta_{12} + \Delta_{21} + \Delta_{22}}{4} \cr 
\frac{\Delta_{11}+\Delta_{12} + \Delta_{21} + \Delta_{22}}{4} & \frac{\Delta_{11}+\Delta_{12} + \Delta_{21} + \Delta_{22}}{4} & \frac{\Delta_{11}+\Delta_{12} + \Delta_{21} + \Delta_{22}}{4} 
\end{pmatrix},
\end{equation}
where $U=U(\frac{\pi}{2},\alpha, \beta)$, and $\alpha, \beta \in \{\frac{\pi}{4},\frac{3\pi}{4},\frac{5\pi}{4},\frac{7\pi}{4}\} $. 

In the case of extending the classical game with two quantum strategies, the general extension matrix is of the form
\begin{equation}\label{general4x4}
\bordermatrix{ & I & iX & U_1 & U_2 \cr I & \Delta_{11} & \Delta_{12} & \Delta_{13} & \Delta_{14} \cr 
iX & \Delta_{21} & \Delta_{22} & \Delta_{23} & \Delta_{24}\cr 
U_1 & \Delta_{31} & \Delta_{32} & \Delta_{33} & \Delta_{34} \cr
U_2 & \Delta_{41} & \Delta_{42} & \Delta_{43} & \Delta_{44}
}.
\end{equation}
 Unitary operators $U_{1}=U_{1}(\theta_1, \alpha_1, \beta_1)$, $U_{2}=U_{2}(\theta_2, \alpha_2, \beta_2)$ and the payoff matrix (\ref{general4x4}) are determined to satisfy the invariance condition of the quantum game with respect to isomorphic transformations of the classical game. As demonstrated in \cite{frackiewicz_permissible_2024}, there are eight distinct classes of $4\times 4$ extensions of the $\Gamma$ game: $A_1, A_2, B_1, C_1, D_1, D_2, E_1$ and $E_2$.

All subsequent expressions for the expansion matrices will employ the specific matrix $\Gamma$ (\ref{classical2x2}), which represents the general form of this game. The first extension classes $A_1$ and $A_2$ are defined by matrices
\begingroup 
\setlength\arraycolsep{4.5pt}
\renewcommand\arraystretch{1.5}
\begin{equation} \label{klasaA}
\medmuskip = 0.2mu
A_1 = \begin{pmatrix}
\Gamma & a_1\Gamma+a_1'\Gamma_3 \\ 
a_1\Gamma+a_1'\Gamma_3 & b_1\Gamma+b_1'\Gamma_3
\end{pmatrix}, \quad 
A_2 = \begin{pmatrix}
\Gamma & a_2\Gamma_2+a_2'\Gamma_1 \\ 
a_2\Gamma_1+a_2'\Gamma_2 & b_2\Gamma_3+b_2'\Gamma
\end{pmatrix},
\end{equation}
\endgroup
where 
\begin{equation}\label{Gamma123}
 \Gamma_1=\begin{pmatrix}
         \Delta_{21} & \Delta_{22}\\
         \Delta_{11} & \Delta_{12}
        \end{pmatrix},
 \hspace{3mm}
\Gamma_2=\begin{pmatrix}
        \Delta_{12} & \Delta_{11}\\
         \Delta_{22} & \Delta_{21}
        \end{pmatrix},
\hspace{3mm}
 \Gamma_3
=\begin{pmatrix}
         \Delta_{22} & \Delta_{21}\\
         \Delta_{12} & \Delta_{11}
        \end{pmatrix}
\end{equation}
are derived from the classical game matrix (\ref{classical2x2}), where the rows, columns, or both have been swapped, $a_i=\cos^{2}\alpha_i$, $a'_i= 1-a_i =\sin^{2}\alpha_i$ oraz $b_i=\cos^{2}2\alpha_i$, $b'_i= 1-b_i =\sin^{2}2\alpha_i$. Other parameters of quantum strategies are defined in \cite{frackiewicz_permissible_2024}, in particular $\theta_1 = 0$ and $\theta_2 =\pi$ for $A_1$ and vice versa for $A_2$. The next class $B_1$ of extensions is defined by the matrix
\begingroup 
\setlength\arraycolsep{4.5pt}
\renewcommand\arraystretch{1.5}
\begin{equation} \label{klasaB}
B_1 = \begin{pmatrix}
\medmuskip = 0.2mu
\Gamma & \frac{\Gamma+\Gamma_1+\Gamma_2+\Gamma_3}{4} \\ 
\frac{\Gamma+\Gamma_1+\Gamma_2+\Gamma_3}{4} & \frac{\Gamma+\Gamma_1+\Gamma_2+\Gamma_3}{4}
\end{pmatrix}. 
\end{equation}
\endgroup
In this case $\theta_1=\theta_2=\frac{\pi}{2}$. Extension of the class $C_1$ is given by the formula
\begingroup 
\setlength\arraycolsep{4.5pt}
\renewcommand\arraystretch{1.5}
\begin{equation}
\medmuskip = 0.2mu
C_1 = \begin{pmatrix} \label{klasaC}
\Gamma & t\frac{\Gamma+\Gamma_3}{2}+t'\frac{\Gamma_1+\Gamma_2}{2} \\ 
t\frac{\Gamma+\Gamma_3}{2}+t'\frac{\Gamma_1+\Gamma_2}{2} & t'^2\Gamma + tt'(\Gamma_1+\Gamma_2) + t^2\Gamma_3
\end{pmatrix}, 
\end{equation}
\endgroup
where $t=\cos^{2}\frac{\theta_1}{2}$, $t'= 1-t =\sin^{2}\frac{\theta_1}{2}$. Remaining parameters are defined in \cite{frackiewicz_permissible_2024}, in particular $\theta_2=\pi-\theta_1$ for the extensions $C$, $D$ and $E$. The next class $D$ can be determined by the matrices:  
\begingroup 
\setlength\arraycolsep{4.5pt}
\renewcommand\arraystretch{1.5}
\begin{equation}\label{klasaD}
\medmuskip = 0.2mu
D_1 = \begin{pmatrix}
\Gamma & t\Gamma+t'\Gamma_2 \\ 
t\Gamma+t'\Gamma_1 & t^2\Gamma + tt'(\Gamma_1+\Gamma_2) + t'^2\Gamma_3
\end{pmatrix}, \quad 
D_2 = \begin{pmatrix}
\Gamma & t\Gamma_3+t'\Gamma_1 \\ 
t\Gamma_3+t'\Gamma_2 & t^2\Gamma + tt'(\Gamma_1+\Gamma_2) + t'^2\Gamma_3
\end{pmatrix}.
\end{equation}
\endgroup
The last class $E$ is determined by the matrices
\begingroup 
\setlength\arraycolsep{4.5pt}
\renewcommand\arraystretch{1.5}
\begin{equation}
\medmuskip = 0.2mu
E_1 = \begin{pmatrix}\label{klasaE}
\Gamma & t\Gamma+t'\Gamma_1 \\ 
t\Gamma+t'\Gamma_2 & t^2\Gamma + tt'(\Gamma_1+\Gamma_2) + t'^2\Gamma_3
\end{pmatrix}, \quad
E_2 = \begin{pmatrix}
\Gamma & t\Gamma_3+t'\Gamma_2 \\ 
t\Gamma_3+t'\Gamma_1 & t^2\Gamma + tt'(\Gamma_1+\Gamma_2) + t'^2\Gamma_3
\end{pmatrix}.
\end{equation}
\endgroup
As the attentive reader may have noticed, for the above extensions we have not given all the values of the $\theta_i, \alpha_i, \beta_i$ parameters of the quantum strategies $U_1$ and $U_2$, as they do not affect the payoff matrices. The missing parameters are necessary to prepare the corresponding quantum gates for the game implementation, which is outside the scope of the current work. They can be found in Table 1 of the paper \cite{frackiewicz_permissible_2024}.
%%%%%%%%%%%%%%%%%%%%%%%%%%%%%%%%%%%%%%%%%%%%%%%%%%%%%%%%%%%%%%%%%%%%%%%%%%%%%%%%%%%%%
%%%%%%%%%%%%%%%%%%%%%%%%%%%%%%%%%%%%%%%%%%%%%%%%%%%%%%%%%%%%%%%%%%%%%%%%%%%%%%%%%%%%%
\section{Attributes of the package}
The package QEGS has been developed to facilitate the calculations of classical game properties and their quantum extensions. In the package, we build on the EWL scheme of quantum extensions of classical games. Using the notation and formulas used to define quantum extensions, we created the corresponding functions in the Mathematica package.

The package facilitates the examination of quantum extensions of classical games defined by $2\times 2$ payoff matrices. Additionally, it enables the investigation of properties of any bimatrix games, inclusive of quantum extensions, while limited to a numerical representation of a payoff matrix.

Using this package, users can, for example, investigate the presence of NE in pure strategies for an extension of a given classical game and, most importantly, depending on the parameters of the extension scheme. This provides an opportunity to evaluate the scenarios wherein quantum extensions might offer advantages over classical strategies. 

It is important to stress that the package is not limited to quantum extensions of classical games. It can also be used to investigate the properties of classical games. For classical games, one can investigate the existence of pure strategy NE depending on the payoffs of a considered game. Moreover, the package allows for determining dominating as well as maximin strategies, both for quantum extensions and classical games. 

The package offers features that assist in researching quantum extensions of classical games, including:
\begin{enumerate}[label=(\alph*)]
    \item construct $\Gamma$ matrices following (\ref{Gamma123}) based on the arbitrary $2\times2$ classical game given by (\ref{classical2x2}),
    \item generate an explicit form of permissible quantum extensions $A_0$, $B_0$ or $C_0$ of a classical game by employing one unitary strategy,
    \item generate an explicit form of permissible quantum extensions $A_1$, $A2$, $B_1$, $C_1$, $D_1$, $D_2$, $E_1$ or $E_2$ of a classical game by employing two unitary strategies,
    \item calculate and highlight NE in pure strategies for numerical bimatrices,
    \item given a numerical matrix with a single continuous parameter $x$, generate a matrix with highlighted NE in pure strategies where the parameter is controlled be a dynamic slider, 
    \item find the maximin strategies in an input bimatrix of a classical game, 
    \item given a numerical matrix with a single continuous parameter $x$, generate the matrix with highlighted maximin strategy for both row and column player where the parameter is controlled be a dynamic slider, 
    \item indicate dominated strategies for both players in an input bimatrix, 
    \item given a numerical matrix with a single continuous parameter $x$, generate the matrix with highlighted dominated strategy for both  players where the parameter is controlled be a dynamic slider, 
    \item generate a PDF report summarizing features and extensions of an input game bimatrix. 
\end{enumerate}

A detailed description of all the features available in the package can be found in  Table \ref{tab:functions} of \ref{appendix_functions}. Moreover, warnings or errors that may be generated by functions of the package are listed in Table \ref{tab:warnings} of \ref{appendix_warnings}.

\section{Examples of use}

The Wolfram Language script QEGS is stored in the file \texttt{QEGS.wl} and is accompanied by two test notebooks \texttt{Extensions.nb} and \texttt{Solver.nb}. In the current section, we will show how to use notebooks and test the functionalities of the package. The two notebooks are organized in such a way that the first \texttt{Extensions.nb} is dedicated to studying the properties of quantum extensions of classical games (functionalities (a) - (i)), and the second \texttt{Solver.nb} is used to investigate properties of classical games (functionalities (d) - (j)).
\subsection{Quantum extensions of classical games}
To test properties of quantum extensions of classical games, an example notebook has been prepared, namely \texttt{Extensions.nb}. The test notebook is initially used by supplying sample inputs.

\begin{verbatim}
MatSym = {{{s1, s1}, {s2, s3}}, {{s3, s2}, {s4, s4}}};
MatSymNum = MatSym /. {s1 -> 3, s2 -> 0, s3 -> 5, s4 -> 1};
MatEx1 = {{{3, 1}, {2, 3}, {2, 0}}, {{-100, 1}, {-100, 2}, {3, 3}}};
\end{verbatim}
The first definition of \texttt{MatSym} 
allows a user to define a symbolic game 
matrix in Mathematica syntax. Whenever it
is necessary to use numerical values 
instead of symbolic ones, it is possible 
to substitute chosen values, which is 
denoted by a definition of 
\texttt{MatSymNum}. Additionally, it is 
possible to test numerical payoff matrices 
by defining them directly. Note that
\texttt{MatEx1} is a game 
defined in Example \ref{Example_3} and given by  
Eq. (\ref{game_2_3}) in which players have different numbers of strategies.

\begin{figure}[h!]
\centering
\includegraphics[width=0.9\linewidth]{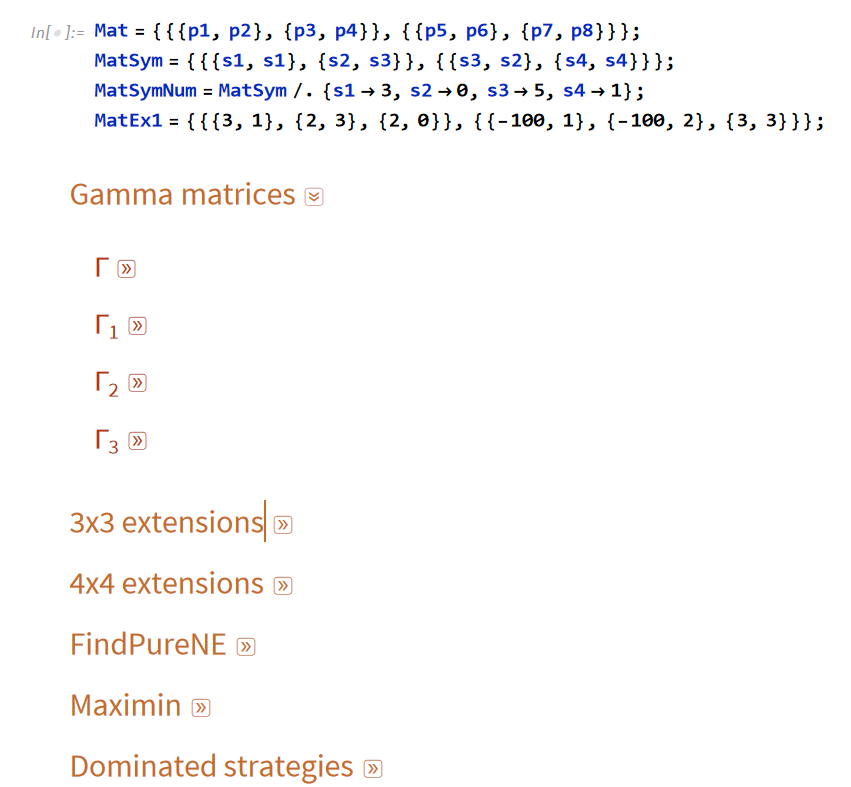}
\caption{The test notebook \texttt{Extensions.nb} structure}\label{Ext_fig_1}
\end{figure}

The test notebook \texttt{Extensions.nb} is organized in the 
way presented in Fig. \ref{Ext_fig_1} where 
subsequent calculations are split into 
sections.

\begin{figure}[h!]
\centering
\includegraphics[width=0.4\linewidth]{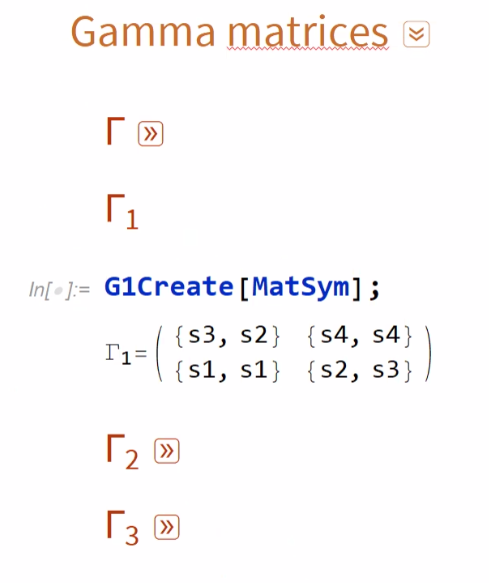}
\caption{$\Gamma_1$ matrix (Eq. (\ref{Gamma123})) calculations for a symbolic input payoff matrix}\label{Ext_fig_2}
\end{figure}

The example notebook begins by computing the explicit form of $\Gamma$ matrices, as indicated by equations (\ref{classical2x2}) and (\ref{Gamma123}). A package user is
required to insert a payoff matrix name as 
an input to a corresponding $\Gamma$ function. It can be applied to both symbolic and numerical matrices. An example use for $\Gamma_1$ is presented in Fig. \ref{Ext_fig_2}.

Similar syntax and output are obtained by the use of further $\Gamma$'s functions, i.e. $\Gamma$, $\Gamma_2$, and $\Gamma_3$.

\begin{figure}[h!]
\centering
\includegraphics[width=1\linewidth]{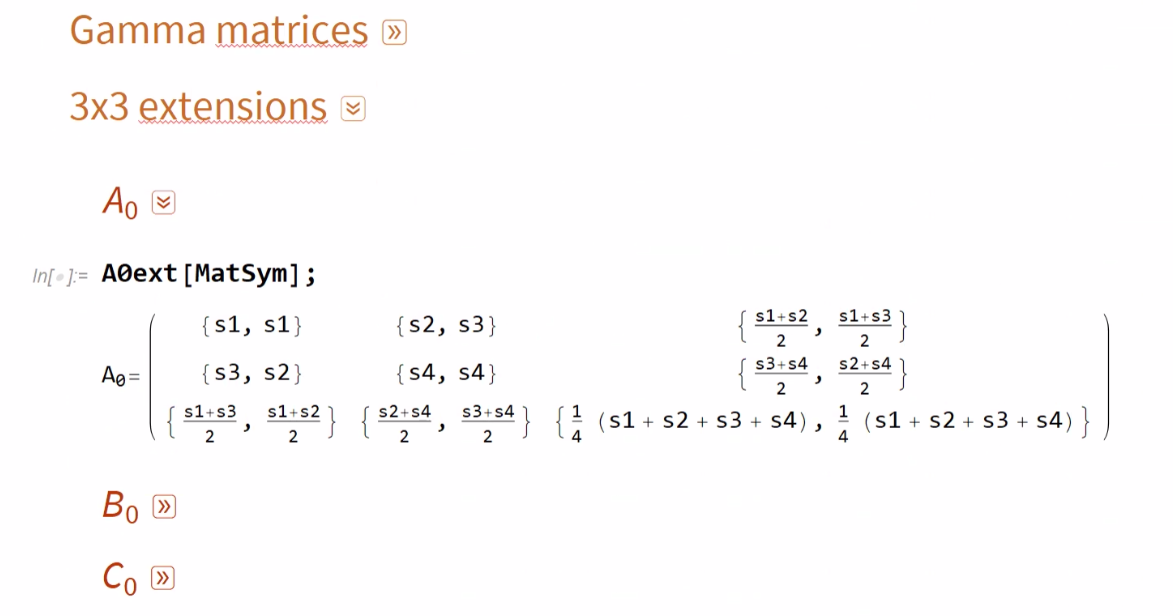}
\caption{Section of the test notebook to investigate extensions of a 
classical game with the use of one
unitary transformation}\label{Ext_fig_3}
\end{figure}

The next section in \texttt{Extensions.nb} 
can be used to test quantum extensions $A_0$, $B_0$ and $C_0$ of classical games obtained by applying a single unitary transformation, according to formulas (\ref{A0}), (\ref{B0}), and (\ref{C_0}). 
The resulting matrices are $3\times 3$ dimensional. The subsequent categories of the extensions can be examined thoroughly in the related subsections, as illustrated in Fig. \ref{Ext_fig_3}. The example extension $A_0$ (Eq. (\ref{A0})) is shown as a result produced by the function.

\begin{figure}[h!]
\centering
\includegraphics[width=0.6\linewidth]{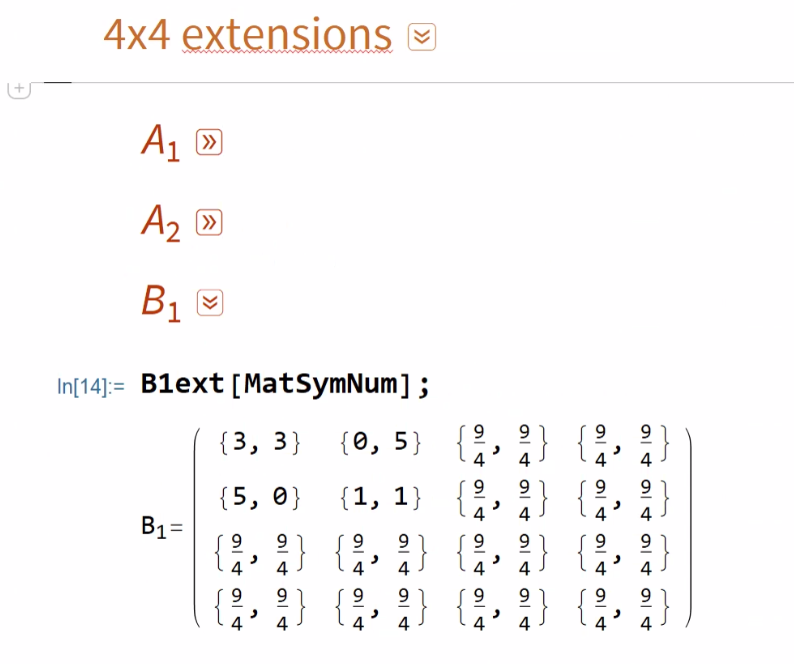}
\caption{Section of a test notebook to investigate $4\times 4$ extensions of a classical game. An example extension, namely $B_1$ has been calculated within this section.}\label{Ext_fig_4}
\end{figure}

An examination of the characteristics of classical games extended by two unitary transformations is included in the $4\times 4$ extensions part of the test notebook, as depicted in Fig. \ref{Ext_fig_4}. The example extension $B_1$ (Eq. (\ref{klasaB})) is shown as a result.

\begin{figure}[h!]
\centering
\includegraphics[width=1.3\linewidth]{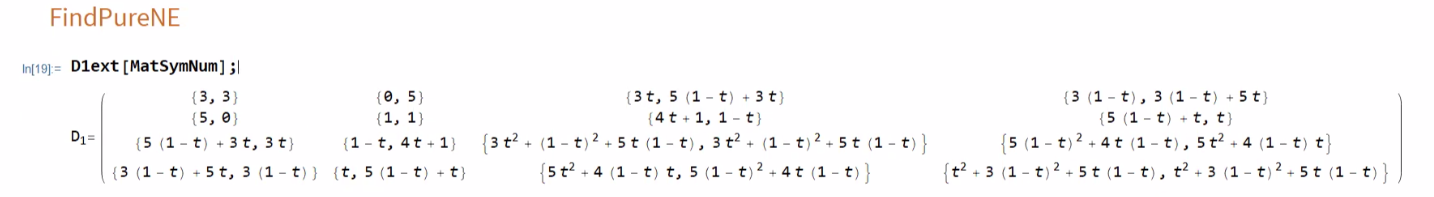}
\caption{Section of the test notebook to investigate pure NE of a given quantum extension of a classical game, namely $D_1$ in the presented example.}\label{Ext_fig_6a}
\end{figure}

\begin{figure}[h!]
\centering
\includegraphics[width=0.9\linewidth]{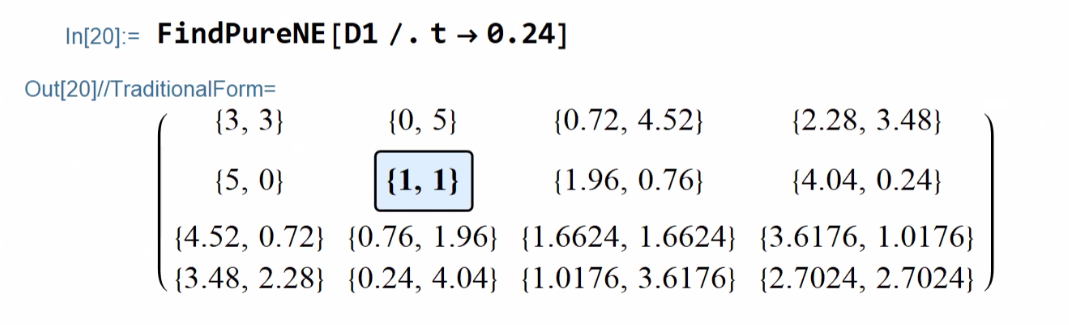}
\caption{A pure NE of $D_1$ extension of Prisoner dilemma with $t=0.24$.  }\label{Ext_fig_6b}
\end{figure}

Additional functions allow for the examination of NE existence within corresponding extensions when provided as an input argument. One can work with \texttt{FindPureNE} function by 
introducing a parametrized quantum extension of a given 
classical game and then setting the parameter to investigate 
the existence of NE related to changes of the 
parameter values. Figures \ref{Ext_fig_6a} and \ref{Ext_fig_6b} demonstrate the example. To determine a NE payoff and the strategy profile associated with the $D_1$ extension (see Eq. (\ref{klasaD})), a matrix is employed where the parameter $t$ is set to 0.24. The NE achieved is represented by a blue rectangle and pertains to the strategy profile (2,2), yielding a payoff of 1 for each player.

\begin{figure}[h!]
\centering
\includegraphics[scale=0.5]{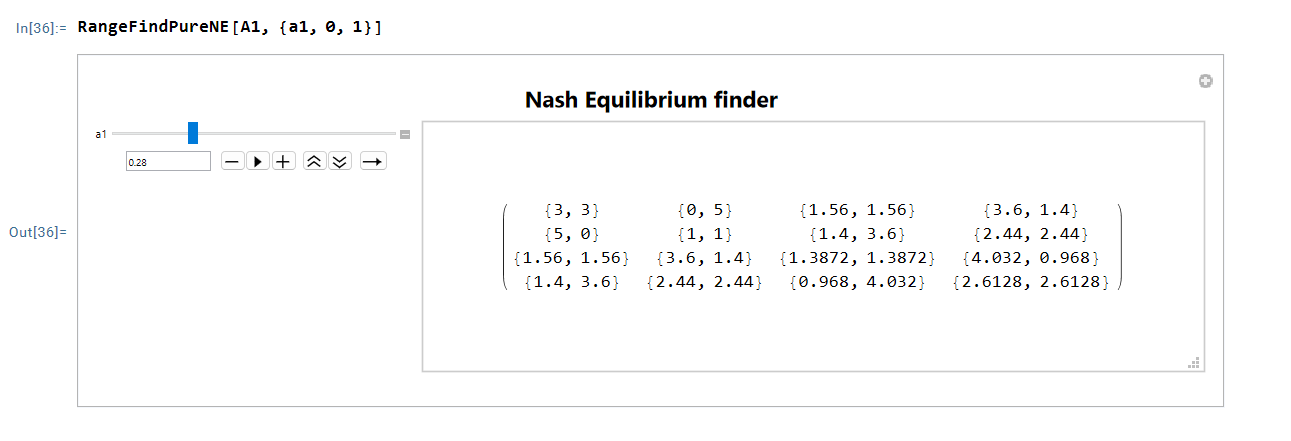}
\includegraphics[scale=0.5]{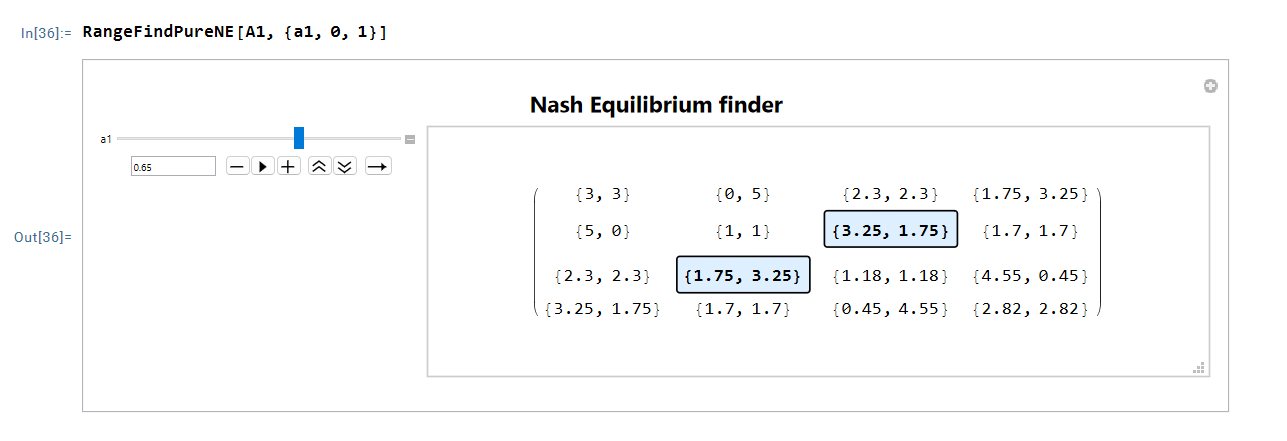}
    \caption{Pure NE finder allowing for dynamic change of a quantum extension parameter.} \label{Ext_fig_8}
\end{figure}

\begin{figure}[h!]
\centering
\includegraphics[scale=0.5]{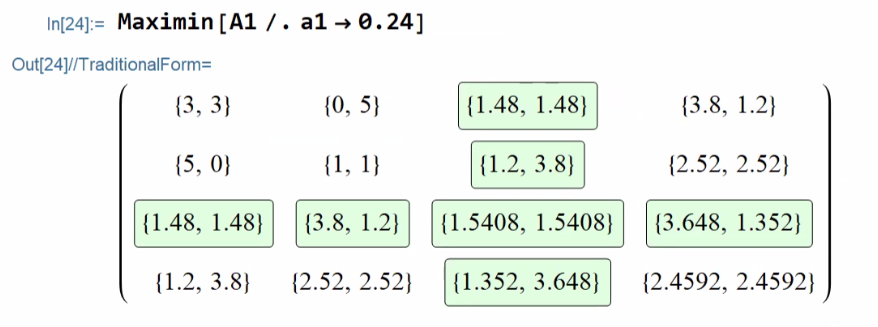}
\caption{Section of a test notebook to investigate maximin strategies of a given quantum extension of a classical game with a set value of its parameter, $A_1$ in the presented example.}\label{Ext_fig_7a}
\end{figure}

\begin{figure}[h!]
\centering
\includegraphics[scale=0.5]{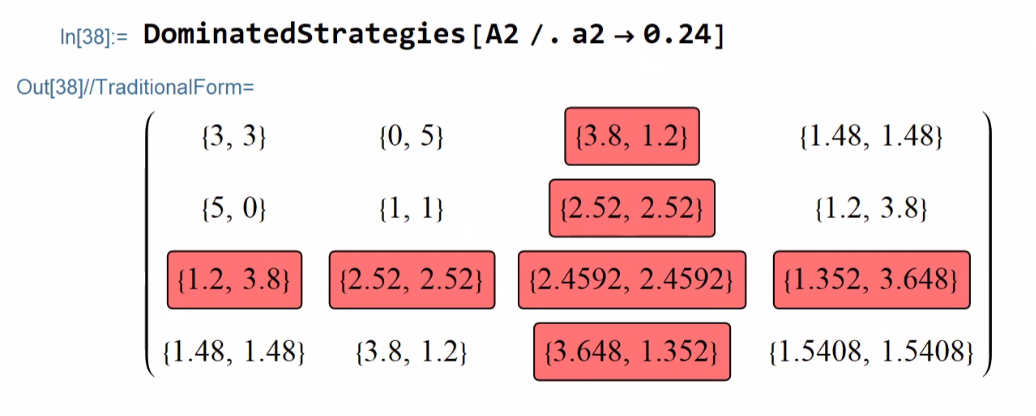}
\caption{Section of a test notebook to investigate dominated strategies of a given quantum extension of a classical game with a set value of its parameter, $A_2$ in the presented example.}\label{Ext_fig_8a}
\end{figure}

Likewise, \texttt{Maximin} and \texttt{DominatedStrategies} provide a means to explore the characteristics of a selected quantum extension when used as an input argument, as illustrated in Figs. \ref{Ext_fig_7a} and \ref{Ext_fig_8a}, respectively. Adjusting a parameter associated with a specific extension allows one to observe how maximin and dominated strategies shift with changes in the parameter. The strategies are highlighted in green for \texttt{Maximin} and in red for \texttt{DominatedStrategies}.

\begin{figure}[h!]
\centering
\includegraphics[scale=0.5]{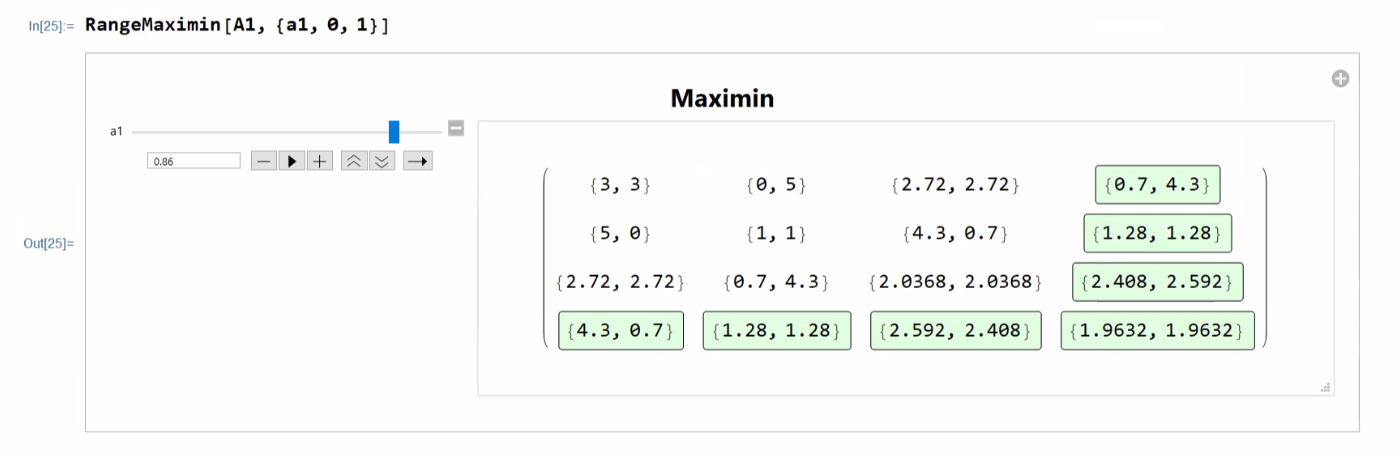}
\caption{Section of a test notebook to dynamically investigate maximin strategies of a given quantum extension of a classical game, $A_1$ in the presented example.}\label{Ext_fig_7b}
\end{figure}

\begin{figure}[h!]
\centering
\includegraphics[scale=0.5]{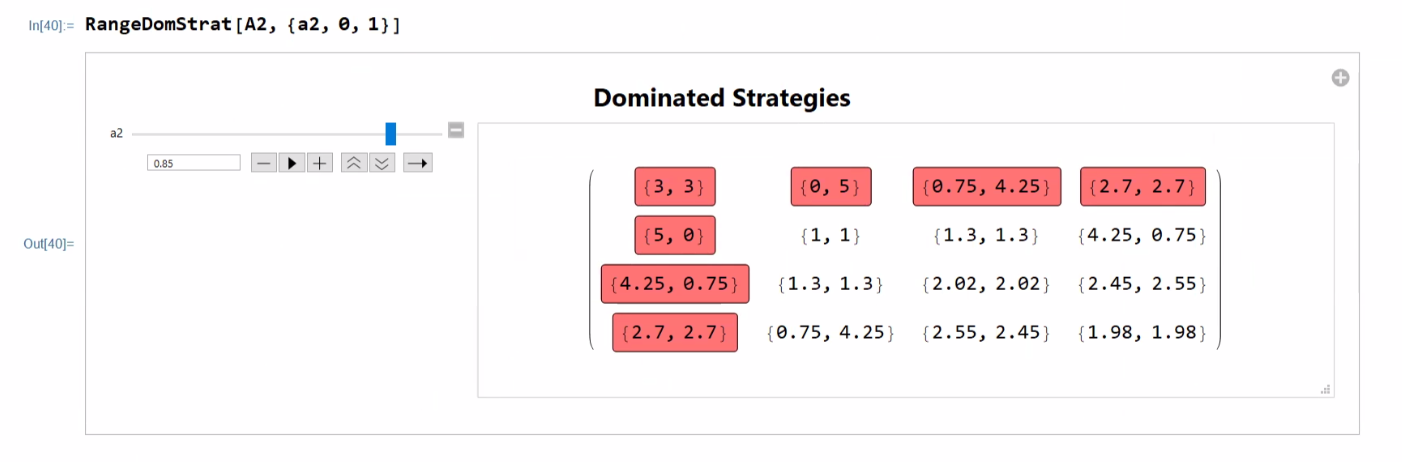}
\caption{Section of a test notebook to dynamically investigate dominated strategies of a given quantum extension of a classical game, $A_2$ in the presented example.}\label{Ext_fig_8b}
\end{figure}

The features designed to analyze the dependency of NE on the parameterization of a quantum game extension, along with variations in maximin and dominated strategies, are implemented using Mathematica's `Manipulate` function.

A slider can be used to adjust a parameter value dynamically within a specified range. Additionally, a specific value can be entered manually without employing the slider. An illustration of this is shown in Fig. \ref{Ext_fig_8} in the context of examining NE. Two panels vary in terms of the extension parameter value. Notably, in the upper panel where $a_1=0.28$, a pure NE is absent. By contrast, when $a_1=0.65$, there are two NE found for the strategy profiles (2,3) and (3,2). The resulting payoffs can be readily obtained from the function's output.

Additional dynamic analytical options for delving into the further properties of quantum extensions have also been made available for \texttt{Maximin} and \texttt{DominatedStrategies}, as depicted in Figures \ref{Ext_fig_7b} and \ref{Ext_fig_8b}, respectively.

\subsection{Game solver}
Certain functions mentioned above exhibit characteristics that are not directly linked to the analysis of quantum extensions, and they can be utilized to explore properties of classical games in their general form.

An additional test notebook, named \texttt{Solver.nb}, is offered for users to investigate various aspects of classical games. This notebook can be used for various types of classical games, providing flexibility for a payoff matrix that need not be square-shaped.

\begin{figure}[h!]
\centering
\includegraphics[scale=0.7]{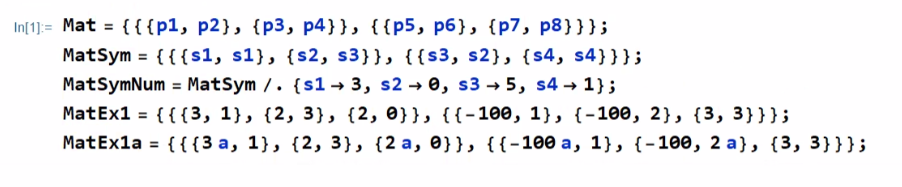}
\caption{The test notebook \texttt{Solver.nb} example of input arguments i.e. classical games payoff matrices.}\label{GS_fig_1}
\end{figure}

Figure \ref{GS_fig_1} showcases different types of input games for analysis using \texttt{Solver.nb}. It highlights the process of representing a classical game symbolically and demonstrates how to parameterize a payoff matrix or replace parameters with specific numbers.

Be aware that \texttt{MatEx1}, as shown by the payoff matrix in Example \ref{Example_1}, is also suitable for examination using this test notebook. Additionally, it allows for parameterization in accordance with Fig. \ref{GS_fig_1}.

\begin{figure}[h!]
\centering
\includegraphics[scale=0.5]{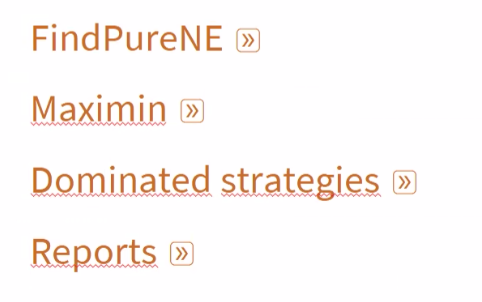}
\caption{The test notebook \texttt{Solver.nb} functions applicable to input payoff matrices.}\label{GS_fig_2}
\end{figure}

The following functions are available
in the test notebook \texttt{Solver.nb} (Fig. \ref{GS_fig_2}):
\begin{itemize}
    \item \texttt{FindPureNE}
    \item \texttt{Maximin}
    \item \texttt{DominatedStrategies}
    \item \texttt{Reports}
\end{itemize}

\begin{figure}[h!]
\centering
\includegraphics[scale=0.3]{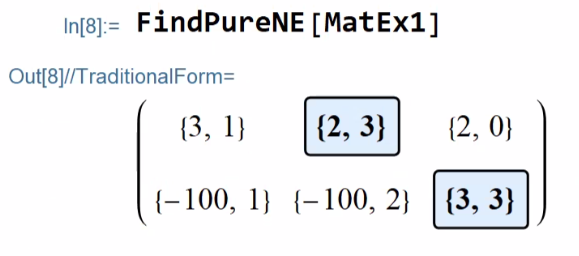}
\includegraphics[scale=0.5]{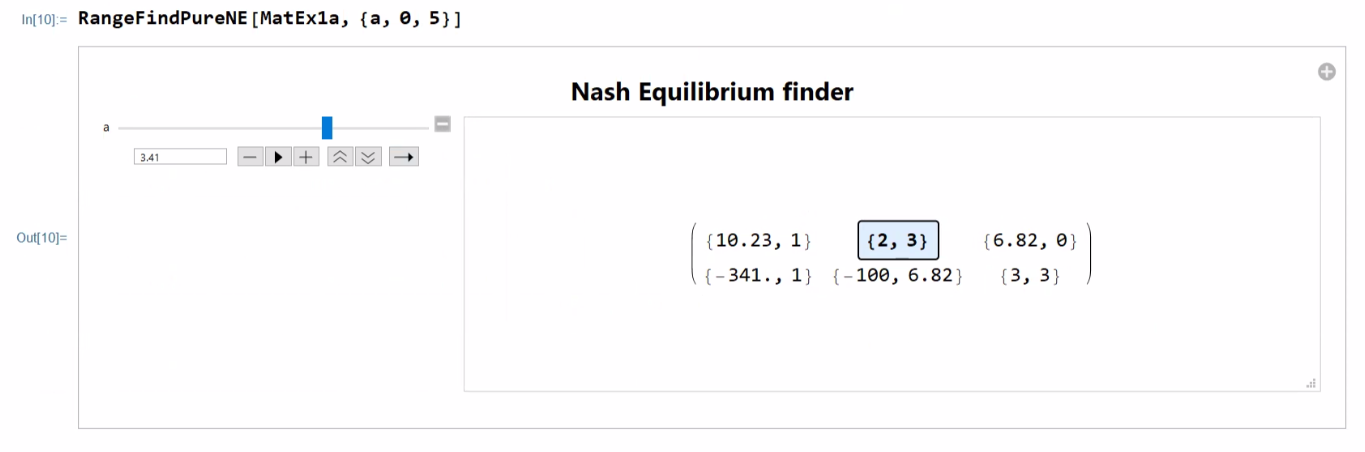}
\caption{Pure NE finder. In the below panel a dynamic change of an input game parameter is possible with a slider in order to investigate existence of NE.} \label{GS_fig_3}
\end{figure}

Using the \texttt{FindPureNE} function, one can determine pure NE for the explicit form of a game. Additionally, the \texttt{RangeFindPureNE} function allows for dynamic analysis of the presence of pure NE in games with parametrized payoffs. Figure \ref{GS_fig_3} illustrates an example employing both functions.

\begin{figure}[ht!]
\centering
\includegraphics[scale=0.3]{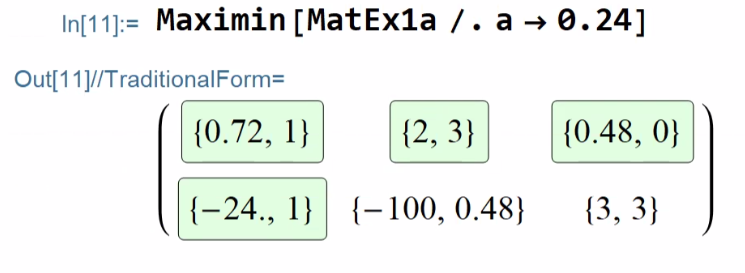}
\includegraphics[scale=0.5]{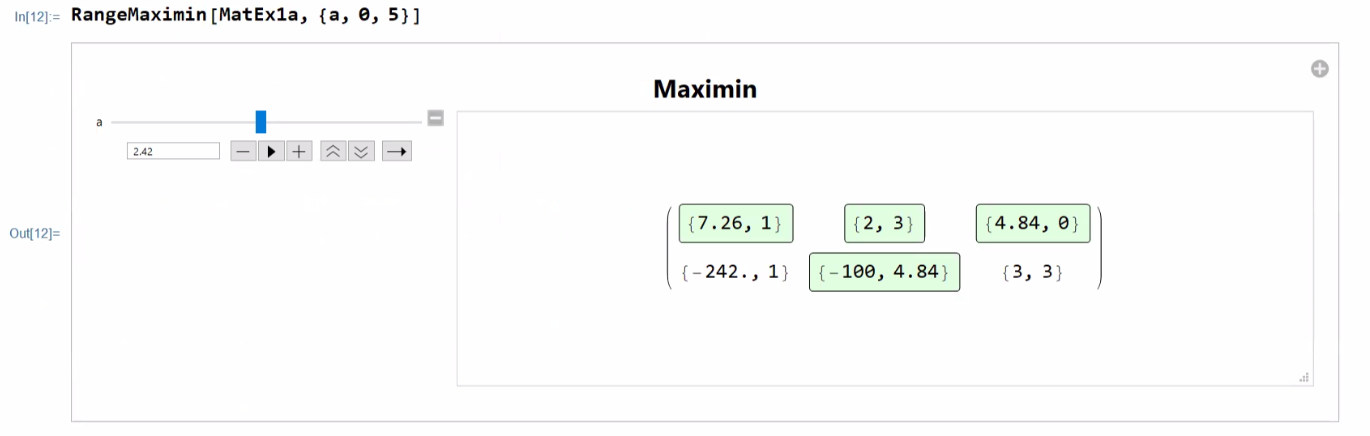}
\caption{Maximin strategies finder.  In the lower panel, a slider allows dynamic alteration of an input game parameter to explore maximin strategies.} \label{GS_fig_4}
\end{figure}

\begin{figure}[ht!]
\centering
\includegraphics[scale=0.3]{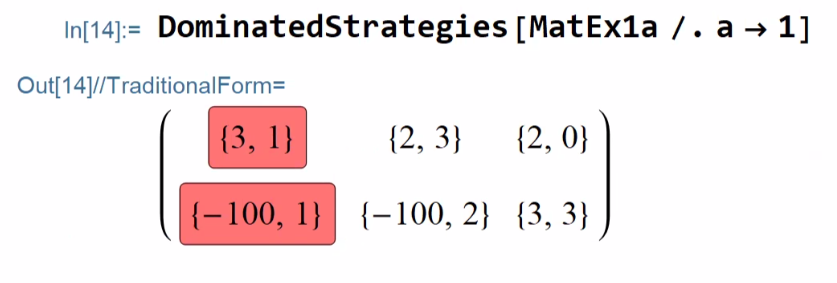}
\includegraphics[scale=0.5]{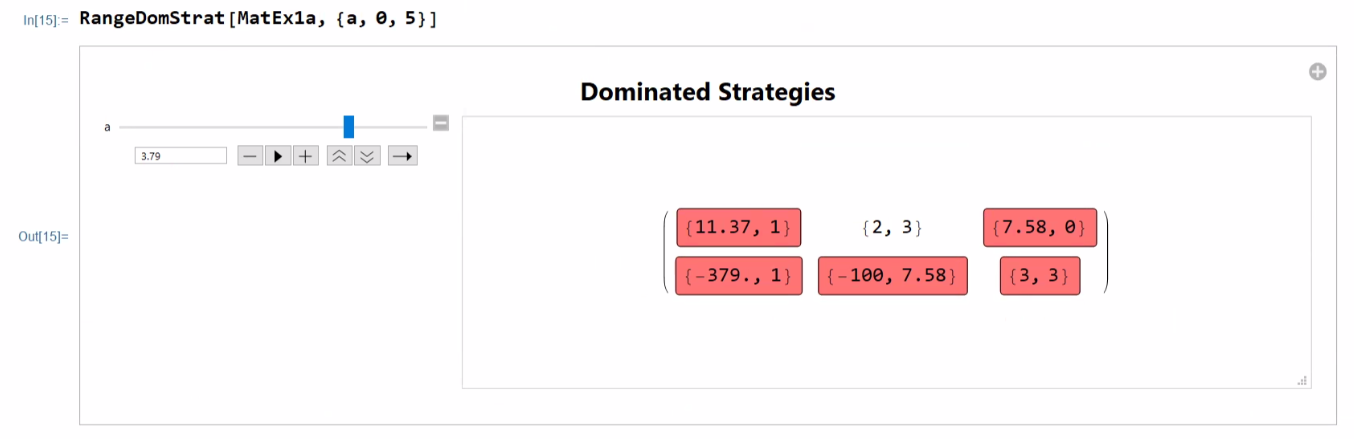}
\caption{Dominated strategies finder. In the lower panel a dynamic change of an input
game parameter is possible with a slider in order to investigate dominated strategies.} \label{GS_fig_5}
\end{figure}

In a similar manner, maximin and dominated strategies for any particular classical game can be examined. Utilizing a slider facilitates the analysis of their presence by varying a parameter within a specified range. This analysis is illustrated in Figs. \ref{GS_fig_4} and \ref{GS_fig_5} for maximin and dominated strategies, respectively.

\texttt{GenerateReport} function is available in the test game solver notebook. The test notebook includes four examples that illustrate different output options based on whether the input matrix is numerical or symbolic. It is assumed that the input matrix is $2\times 2$ for extensions analysis.
For $2\times 3$ numerical matrices a simplified output is 
also generated. Depending on an input game, possible outputs of \texttt{GenerateReport} function and corresponding file names are listed below. Calling the function \texttt{GenerateReport[Mat,name]} results in the following files with corresponding names:
\begin{itemize}
    \item numerical, $2 \times 2$ input matrix game {\bf Mat}: \texttt{Report\_name.pdf}
    \item numerical input matrix game {\bf Mat}: \texttt{Report\_name\_extensions.pdf}
    \item symbolic, , $2 \times 2$ input matrix game {\bf Mat}: \texttt{Report\_name\_properties.pdf}
    \item symbolic input matrix game {\bf Mat}: No report
\end{itemize}

In detail, the properties of investigated games covered in the corresponding reports are listed below: 
\begin{enumerate}
    \item Output for $2\times 2$ numerical input matrix \label{item_report_1}
    \begin{itemize}
        \item Quantum extensions of the input game with one unitary strategy
        \item Quantum extensions of the input game with two unitary strategies
        \item NE of the input game
        \item Both players' maximin strategies of the input game
        \item Both players' dominated strategies of the input game
    \end{itemize}
    \item Output for $2\times 2$ general input matrix \label{item_report_2}
    \begin{itemize}
        \item Quantum extensions of the input game with one unitary strategy
        \item Quantum extensions of the input game with two unitary strategies
    \end{itemize}
    \item Output for $n\times m$ numerical input matrix \label{item_report_3}
    \begin{itemize}
        \item NE of the input game
        \item Both players' maximin strategies of the input game
        \item Both players' dominated strategies of the input game
    \end{itemize}
     \item Output for $n\times m$ symbolic input matrix \label{item_report_4}
     \begin{itemize}
        \item no output
    \end{itemize}
\end{enumerate}

The output file name is constructed as 
presented in Fig. \ref{Fig_Flowchart_1}. 
Furthermore, the process of creating reports 
covering the results of the analysis, depending on an input game matrix, and 
processed by the \texttt{GenerateReport} 
function is presented in Fig. \ref{Fig_Flowchart_2}.

\section{QEGS GitHub repository}
The QEGS package is available in a 
dedicated GitHub repository under the 
following link:  \href{https://github.com/k-grzanka/QEGS}{https://github.com/k-grzanka/QEGS}

The package is accompanied by two test notebooks
\texttt{Extensions.nb} and \texttt{Solver.nb}, 
also available in the Github repository. The files
are shared under the GNU General Public License version 3.

\section{QEGS - future horizons}
For future outlook, we will continue the 
development of the QEGS package with a 
collaborative and iterative mindset. 
Based on possible feedback on the 
current version of the QEGS, we 
plan to update the package if necessary 
and develop additional features.

In our future work, we would like to 
take into account the following aspects:
\begin{enumerate}
    \item iterative methods for finding dominated strategies;
    \item development of new functions with sliders for more than one parameter;
    \item investigation of NE in mixed strategies.
\end{enumerate}
These advanced features 
will enable package users to optimize their decision-making processes while utilizing game theory and improve the overall functionality of our package.

In our future work, we would like to take into account such aspects as sustainability by optimizing resource usage, especially with the growing size of game matrices, and fostering a culture of continuous learning and innovation within the team. By embracing these elements, we would like to create an updated version of the QEGS package that not only meets current demands but also anticipates future trends and challenges in quantum game theory.

\section{Summary}
Quantum game theory is complex due to its 
interdisciplinary nature, combining physics, 
computer science, mathematics, and economics.
This complexity poses a challenge for new 
researchers. With advancements in quantum 
computing, there is a growing focus on its 
security implications and computational 
benefits. Consequently, ways of building 
strategies for players following classical game theory are reaching new 
horizons via e.g. using unitary strategies 
according to EWL. 

Yet, due to the complexity 
of quantum extensions properties, a tool is 
needed to support analytical calculations. 
That is the rationale to develop a dedicated 
Mathematica package to automatize such 
analysis. QEGS package allows a user to 
investigate quantum extensions of a given 
classical game as well as properties of an 
input game itself. One can test various forms
of input classical games and improve their 
understanding of game theory aspects without 
a necessity of tedious analytical 
calculations. 

\section{Declaration of generative AI and AI-assisted technologies in the writing process}
During the preparation of this work, the authors used Writefull integrated with Overleaf and ChatGPT in order to improve the readability and language of the manuscript. After using these tools, the authors reviewed and edited the content as needed and take full responsibility for the content of the published article.

\newpage
 \bibliographystyle{elsarticle-num} 
 \bibliography{references}

\appendix

\section{Functions of the package}\label{appendix_functions}

To keep the notation concise, we introduce a definition of a matrix in Mathematica syntax to be
processed by the subsequent
functions defined in the package:
\begin{equation}
    \operatorname{Mat}=\{\{\{p1,p2\},\{p3,p4\}\},\{\{p5,p6\},\{p7,p8\}\}\}. 
\end{equation}
The corresponding payoffs bimatrix in a traditional form is the following:
\begin{equation}
\operatorname{Mat} = \begin{pmatrix}
(p1,p2) & (p3,p4) \\ 
(p5,p6) & (p7,p8) 
\end{pmatrix}. \label{Mat}
\end{equation}

In addition, we also define an input list to three functions, namely, RangeFindPureNE, RangeMaximin, RangeDomStrat. 
\begin{equation}
    \operatorname{List}=\{x, x_{\rm{min}}, x_{\rm{max}}\}.
\end{equation}
These functions exploit Mathematica's  Manipulate which generates a version of Mat with additional controls to enable interactive manipulation of $x$ within the range $\{x, x_{\rm{min}}, x_{\rm{max}}\}$.

The table below captures all the functions defined in the package along 
with their description, examples of use, and the anticipated output.

The 'Output' column describes a reference name for a particular variable being an output of a particular package function.

\newpage 

{\small{
\begin{longtable}{|p{3.5cm}|p{8.2cm}|c|}
\hline
\textbf{Function} & \multicolumn{1}{c|}{\textbf{Description}}& \textbf{Output} \\ 
\hline
\hline
GCreate & \multicolumn{1}{c|}{GCreate[Mat,quiet*]} &  G \\
\cline{2-2}
 & Create a classical game payoff matrix $\Gamma$ (Eq. (\ref{classical2x2})) given by a bimatrix in a form of \textbf{Mat}. Argument 'quiet' by default is False and prints output matrix in a traditional form & \\
\hline
 G1Create & \multicolumn{1}{c|}{G1Create[Mat,quiet*]} & G1 \\
\cline{2-2}
 & Generate payoff bimatrix $\Gamma_1$, given by Eq. (\ref{Gamma123}), by swapping rows of the basic game   &  \\
\hline
 G2Create & \multicolumn{1}{c|}{G2Create[Mat,quiet*]} & G2 \\
\cline{2-2}
 & Generate payoff bimatrix $\Gamma_2$, given by Eq. (\ref{Gamma123}), by swapping columns of the basic game &   \\
\hline
 G3Create & \multicolumn{1}{c|}{G3Create[Mat,quiet*]} &  G3 \\
\cline{2-2}
 & Generate payoff bimatrix $\Gamma_3$, given by Eq. (\ref{Gamma123}), by swapping rows and columns of the basic game &  \\
\hline
A0ext& \multicolumn{1}{c|}{A0ext[Mat]} & A0 \\
\cline{2-2}
 & Generate class $A_0$ three-strategy quantum extension for bimatrix game in a form of \textbf{Mat}. It is described by the formula (\ref{A0})  &  \\
 \hline
B0ext & \multicolumn{1}{c|}{B0ext[Mat]} & B0 \\
\cline{2-2}
 & Generate class $B_0$ three-strategy quantum extension for bimatrix game in a form of \textbf{Mat}. It is described by the formula (\ref{B0}) &   \\
\hline
C0ext& \multicolumn{1}{c|}{C0ext[Mat]} &  C0 \\
\cline{2-2}
 & Generate class $C_0$ three-strategy quantum extension for bimatrix game in a form of \textbf{Mat}. It is described by the formula (\ref{C_0}) &   \\
\hline
A1ext & \multicolumn{1}{c|}{A1ext[Mat]} &  A1 \\
 \cline{2-2}
 & Generate class $A_1$ four-strategy quantum extension for bimatrix game in a form of \textbf{Mat}. It is described by the formula (\ref{klasaA}) &  \\
\hline
A2ext & \multicolumn{1}{c|}{A2ext[Mat]}  & A2 \\
 \cline{2-2}
 & Generate class $A_2$ four-strategy quantum extension for bimatrix game in a form of \textbf{Mat}. It is described by the formula (\ref{klasaA}) &  \\
\hline
B1ext & \multicolumn{1}{c|}{B1ext[Mat]} &  B1 \\
 \cline{2-2}
 & Generate Class $B_1$ four-strategy quantum extension for bimatrix game in a form of \textbf{Mat}. It is described by the formula (\ref{klasaB})  & \\
\hline
 C1ext& \multicolumn{1}{c|}{C1ext[Mat]} & C1 \\
 \cline{2-2}
 & Generate Class $C_1$ four-strategy quantum extension for bimatrix game in a form of \textbf{Mat}. It is described by the formula (\ref{klasaC}) &  \\
\hline
 D1ext & \multicolumn{1}{c|}{D1ext[Mat]} & D1 \\
 \cline{2-2}
 & Generate Class $D_1$ four-strategy quantum extension for bimatrix game in a form of \textbf{Mat}. It is described by the formula (\ref{klasaD}) &  \\
\hline
 D2ext& \multicolumn{1}{c|}{D2ext[Mat]} &  D2 \\
 \cline{2-2}
 & Generate Class $D_2$ four-strategy quantum extension for bimatrix game in a form of \textbf{Mat}. It is described by the formula (\ref{klasaD}) &  \\
\hline
 E1ext & \multicolumn{1}{c|}{E1ext[Mat]} &  E1 \\
 \cline{2-2}
 & Generate Class $E_1$ four-strategy quantum extension for bimatrix game in a form of \textbf{Mat}. It is described by the formula (\ref{klasaE}) & \\
\hline
 E2ext & \multicolumn{1}{c|}{E2ext[Mat]} & E2 \\
 \cline{2-2}
 & Generate Class $E_2$ four-strategy quantum extension for bimatrix game in a form of \textbf{Mat}. It is described by the formula (\ref{klasaE}) &  \\
\hline
 FindPureNE & \multicolumn{1}{c|}{FindPureNE[Mat]} & NEmatrix \\
 \cline{2-2}
 & Highlight NE in a numerical bimatrix &  \\
\hline
RangeFindPureNE & \multicolumn{1}{c|}{RangeFindPureNE[Mat,List]} & \\
 \cline{2-2}
 & Generate the matrix with highlighted NE of a given bimatrix, where payoffs can be expressed in terms of a continuous parameter $x$ defined in argument \textbf{List} in range from $x_{\rm min}$ to $x_{\rm max}$. The parameter value can be controlled with a dynamic slider &  \\
\hline
 Maximin & \multicolumn{1}{c|}{Maximin[Mat]} & Maximat \\
 \cline{2-2}
 & Highlight both players’ maximin strategies in the input bimatrix & \\
\hline
 RangeMaximin & \multicolumn{1}{c|}{RangeMaximin[Mat,List]} & \\
 \cline{2-2}
 & Generate the matrix with highlighted maximin strategy for both row and column player allowing for the $x$ parameter change in specified range from $x_{\rm min}$ to $x_{\rm max}$. The parameter value can be controlled with a dynamic slider &  \\
\hline
 DominatedStrategies & \multicolumn{1}{c|}{DominatedStrategies[Mat]} & DomMat \\
 \cline{2-2}
 & Highlight dominated strategies for both players in the input bimatrix. &  \\
\hline
 RangeDomStrat & \multicolumn{1}{c|}{RangeDomStrat[Mat,List]} &  \\
  \cline{2-2}
  & Generate the matrix with highlighted dominated strategies for both players allowing for the $x$ parameter change in specified range from $x_{\rm min}$ to $x_{\rm max}$. The parameter value can be controlled with a dynamic slider & \\
\hline
 GenerateReport & \multicolumn{1}{c|}{GenerateReport[Mat,name]} & PDF file\\
 \cline{2-2}
 & Make a PDF report Report\_{name}\_X.pdf summarizing features and (if possible) extensions of the input game bimatrix &  \\
\hline
\caption{List of all functions defined in the package QEGS. Argument denoted by * is optional.}
\label{tab:functions}
\end{longtable}}}

\section{Warnings and errors raised in the package}\label{appendix_warnings}

The next table describes the warnings and 
possible errors that a user
may encounter while working with the package. 

{\small{
\begin{longtable}{|p{1.8cm}|p{1.8cm}p{3.3cm}|p{6.1cm}|}
\hline
\textbf{Warning/ \hspace{0.5cm} Error} & \multicolumn{2}{c|}{\textbf{Functions}}   & \multicolumn{1}{c|}{\textbf{Solution}}  \\ 
\hline
\hline
Quiet & GCreate G1Create & G2Create \hspace{0.5cm} G3Create  & Argument \texttt{quiet}* should be True/False. The corresponding $\Gamma$ matrix is created anyway.\\
\hline
% Input & \multicolumn{1}{c|}{G1Create } & G1 \\
 Input &  FindPureNE  Maximin & DominatedStrategies  RangeDomStrat GenerateReport & The input matrix \textbf{Mat} needs to be numerical. \\
\hline
Size & A0ext A1ext A2ext  B0ext B1ext & C0ext C1ext \hspace{0.5cm} D1ext D2ext \hspace{0.5cm} E1ext E2ext \hspace{0.5cm} GenerateReport   & The input matrix \textbf{Mat} needs to be 2$\times 2$ dimension. \\
\hline
Error &  \multicolumn{2}{l|}{GenerateReport}  &The input matrix {\bf{Mat}} needs to be either numerical or $2\times2$, otherwise no report is created. \\
\hline
\caption{ Catalog of warnings and errors associated with the respective functions of the QEGS package\\ }
\label{tab:warnings}
\end{longtable}}}

\section{Flowcharts}

\begin{figure}[ht!]
\centering
\includegraphics[scale=0.58]{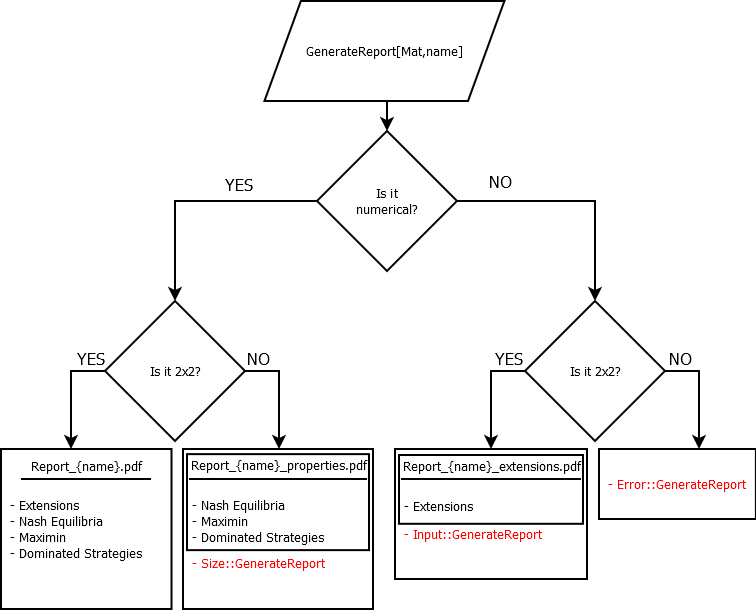}
\caption{Flow of the \texttt{GenerateReport} function with the resulting output files. Warnings that may occur are highlighted in red.} \label{Fig_Flowchart_1}
\end{figure}

\begin{figure}[ht!]
\centering
\includegraphics[angle=90,scale=0.49]{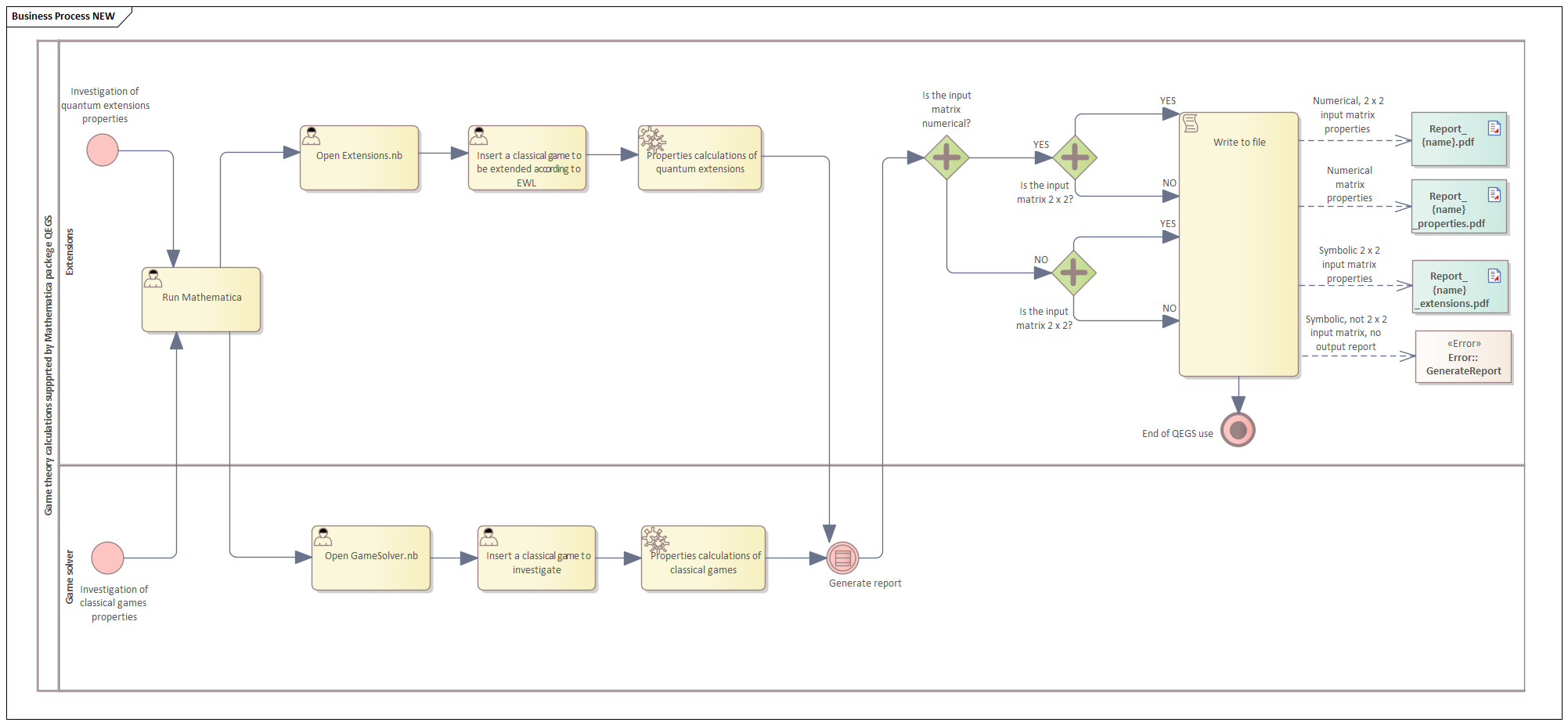}
\caption{Process of calling \texttt{GenerateReport} function.} \label{Fig_Flowchart_2}
\end{figure}

\end{document}